\documentclass{aastex}
\usepackage{emulateapj5}
\usepackage{onecolfloat5}

\input epsf
\def\plotone#1{\centering \leavevmode
\epsfxsize= 1.0\columnwidth \epsfbox{#1}}


\newcommand{\rad}{r}    % comoving radial distance

\newcommand{\da}{d_A}   % comoving angular diameter distance

\def\vecl{{\bf l}}
\def\veck{{\bf k}}

\long\def\comment#1{}

\def\W2{{\cal W}}

\newcommand{\bn}{\hat{\bf n}}

\newcommand{\cmb}{\Theta}

\newcommand{\tot}{{\rm tot}}

\newcommand{\len}{\phi}

\def\be{\begin{equation}}
\def\ee{\end{equation}}
\def\bea{\begin{eqnarray}}
\def\eea{\end{eqnarray}}

\def\cmm2{{\,\rm cm^{-2}}}
\def\cm2{{\,{\rm cm}^2}}
\def\cmm3{{\,{\rm cm}^{-3}}}
\def\gcmm3{{\,{\rm g\,cm^{-3}}}}

\def\fun#1#2{\lower3.6pt\vbox{\baselineskip0pt\lineskip.9pt
  \ialign{$\mathsurround=0pt#1\hfil##\hfil$\crcr#2\crcr\sim\crcr}}}

\def\planck{{\it Planck}}

\def\msun{{\,M_\odot}}
\lefthead{Song, Cooray, Knox $\&$ Zaldarriaga}
\righthead{Far-infrared background --- CMB lensing Cross-Correlation}

\begin{document}
%%% \nopagebreak

%%% \vspace{-\baselineskip}

\twocolumn[%%% Begin front material
\submitted{Submitted to ApJ}

\title{THE FAR-INFRARED BACKGROUND CORRELATION WITH CMB LENSING}
\author{Yong-Seon Song$^{1}$, Asantha Cooray$^{2}$, Lloyd Knox$^{1}$ and Matias Zaldarriaga$^{3}$}
\affil{$^{1}$ Department of Physics, University of California, Davis, CA 95616, USA \\
email: yssong@bubba.ucdavis.edu,lknox@ucdavis.edu}

\affil{$^{2}$ Theoretical Astrophysics, California Institute of Technology, Pasadena, CA 91125, USA\\
email: asante@caltech.edu}
\affil{$^{3}$ Physics Department, New York University, New York , NY 10003, USA\\
Institute for Advanced Study, Einstein Drive, Princeton NJ 08540, USA\\ 	
email: mz3@nyu.edu}

\begin{abstract}
The intervening large--scale structure distorts cosmic microwave
background (CMB) anisotropies via gravitational lensing.  The same
large--scale structure, traced by dusty star--forming galaxies, also
induces anisotropies in the far--infrared background (FIRB).  We
investigate the resulting inter--dependence of the FIRB and CMB with a
halo model for the FIRB.  In particular, we calculate the
cross--correlation between the lensing potential and the FIRB.
The lensing potential can be quadratically estimated from
CMB temperature and/or polarization maps.  We show that the
cross--correlation can be measured with high signal--to--noise
with data from the {\it Planck Surveyor}. We discuss how such a
measurement can be used to understand the nature of FIRB sources and
their relation to the distribution of dark matter.
\end{abstract}

 ]%%% End front material

\section{Introduction}

Dusty star--forming galaxies give rise to a
far--infrared background (FIRB) \citep{puget,fixsen98,dwek,sfd,lagache,guiderdoni98,blain99}.
Correlations in the large--scale structure traced by these
contributing sources lead to correlated fluctuations in the FIRB
\citep{bond86,sw00,hk00,knox01,magliocchetti01}.  At arcminute 
scales and more,
fluctuation power associated with the source distribution can
potentially be detected with Planck and other planned CMB experiments
with channels at frequencies around and above 300 GHz \citep{knox01}.

The same large--scale structure that generates FIRB anisotropy also
generates anisotropy in the CMB in several ways.  These include
modifications due to scattering via free electrons in galaxy clusters,
such as the thermal Sunyaev-Zel'dovich effect \citep{SZ}, and
modifications imposed by the time evolving gravitational field, such
as the integrated Sachs-Wolfe effect \citep{ISW}. The large--scale
structure mass field also deflects CMB photons propagating to us from
the last scattering surface via the gravitational lensing effect
\citep{seljak96}. Since the lensing effect on CMB anisotropies is second
order in temperature fluctuations, it induces non-Gaussian signatures
in the temperature data; the cross--correlation between the lensing effect and other
secondary anisotropies, such as SZ or ISW, contributes to the
temperature bispectrum \citep{GS,CH00,SM99}.

We extend previous discussions on correlations between the CMB and
large--scale structure and consider the cross-correlation of CMB
anisotropies and FIRB fluctuations.  The FIRB contributes
significantly at the high frequency end of certain CMB experiments,
such as the 350 GHz, 545 GHz and 850 GHz channels of the High Frequency
Instrument (HFI) of the Planck Surveyor.  Over this range of
frequencies, and in regions of the sky with low galactic dust
emission, the FIRB stands out as the dominant source of fluctuation
power over a wide range of angular scales \citep{knox01}.  We model
the expected cross-correlation between lensing potentials and the FIRB
though a physically motivated semi-analytical approach involving the
halo model \citep{scherrer,seljak,scocci,cooray02}.

Though the lensing potential-FIRB correlation leads to a bispectrum or
a three-point correlation function in CMB temperature and FIRB, we
suggest a direct cross-correlation between lensing potentials and
FIRB. Our suggestion follows from recent discussions in the literature
on how to reconstruct lensing potentials associated with CMB lensing,
especially through higher order statistics such as quadratic estimates
optimized for the lensing extraction in temperature and polarization
data \citep{zs99,hu011b,hu01,ck02}.  We consider this possibility for
planned missions such as Planck.

Using reasonable assumptions, we show that Planck has sufficient sensitivity
for a detection of the lensing-FIRB correlation.  A detection of
such a correlation would allow us to test how well the sources of the FIRB
trace the large--scale mass distribution.  While semi--analytic
approaches such as the halo model suggest a high correlation, the exact
correlation as measured will allow us to further refine details of
these models and to understand certain physical properties of
contributing FIRB sources.

The layout of the paper is as follows. In Section 2 we present the
halo model for FIRB fluctuations.  In Section 3 we revisit the angular
power spectrum of the FIRB and describe the cross correlation between
the FIRB and a quadratic function of the CMB temperature map.
The quadratic CMB statistic is an estimator for the 
lensing potential.  In section 4 we discuss associated errors and the
extent to which the FIRB-lensing correlation can be detected.
We conclude with a discussion of our results in Section
5. We refer the reader to \citet{hk00} and \citet{knox01} for initial
details on our calculation of correlations in the FIRB.  More details
of the halo approach to large--scale structure are available in the
recent review by Cooray \& Sheth (2002). For a review of our observational
knowledge of the FIRB see \citet{hauser01}.  While we provide a general
derivation of the FIRB-lensing correlation, when illustrating our results
we assume a $\Lambda$CDM cosmological model with
$\Omega_m=0.35,\Omega_\Lambda=0.65,\Omega_b=0.05,h=0.65$. To describe
linear clustering, we use the transfer function given by \citet{eisen}
and normalize fluctuations to COBE \citep{BunWhi97} such that
$\sigma_8=0.86$.

\section{Halo Approach to the FIRB}

As in HK00, we write the FIRB Rayleigh--Jeans temperature ($T_{\rm RJ}
\equiv I_\nu/(2k_B)(c/\nu)^2$) at frequency $\nu$ and in direction
$\hat{n}$ as a sum of the mean and the fluctuation about the mean
\bea
T_{\rm RJ}(\hat{n},\nu)&=&\bar{T}_{\rm RJ}(\nu)+\delta T_{\rm RJ}(\hat{n},\nu) \nonumber \\
&=&{c^2 \over 2 k_B \nu^2}\int
dr\, a(r)\bar{j}(\nu,r) \left[1+\frac{\delta
j(r\hat{n},\nu,r)}{\bar{j}(\nu,r)}\right] \, . \nonumber \\  \eea 
Here, $r$ is the radial distance from us to a redshift of $z$ and
$\bar{j}(\nu,z)$ is the mean emissivity of the dust. 
Note that we have related the excess fluctuation in FIRB temperature
to a fluctuation in the emissivity, $\delta j(r\hat{n},\nu,z)$.

We adopt the HK00 model for the mean emissivity as a function of
redshift and frequency. The key
ingredients of this model are the history of ultra--violet
radiation and dust production.  Both of these are assumed to be
proportional to a star--formation rate, which here and in the HK00
fiducial model we take to be that of \citet{madau}.  By further
assuming the optical properties of the dust \citep{draine84} we
then derive the comoving mean emissivity, $\bar j_\nu(z)$.  The
dust and UV production proportionality constants are chosen so
that the resulting FIRB agrees with inferences from {\it
COBE}/FIRAS data \citep{fixsen98}.

To model fluctuations of the FIRB, we assume that the sources of
this emissivity are galaxies so that 
$\delta j(r\hat{n},\nu,z)/\bar{j}(\nu,z) = \delta_{\rm gal}$ where
$n_{\rm gal} = \bar{n}_{\rm gal}(1+\delta_{\rm gal})$.  
Our modeling differs
from that in HK00 in that we calculate the statistical relation
of the galaxy number density field to the dark matter by making assumptions
about how these galaxies populate dark matter halos; i.e., we use
the `halo approach' to large--scale structure.  

The halo approach to large--scale structure involves a simplified
description of the complex distribution of dark matter as consisting
of a population of dark matter halos. The spatial distribution of any
tracer field of the large--scale structure, such as galaxies, is
described through the relation of the tracer field to that of the dark
matter halo distribution. The halo approach has now been widely
discussed in the literature to understand the clustering of dark
matter, galaxies and baryons \citep{scherrer,scocci,cooray02}. The
necessary inputs for the halo--based calculations come from either
analytical tools, such as the dark matter halo mass function, or from
numerical simulations, such as the distribution of dark matter within
each halo.

The two--point correlation function of the object of interest
in a halo model has two terms; the first is a two-point
correlation function within each dark matter halo and the second
is the two-point correlation function between points which are in
different halos.  The halo distribution is taken to
trace the linear fluctuations with biases which are predictable
from analytical arguments. The time evolution of the correlation
function is naturally determined by this semi--analytic model
since clustering evolution can be related to the evolution of the
halo mass function and any evolution of physical properties
related to halos. As we find below, the halo model has the
advantage that one does not need to specify a priori quantities
such as the bias and evolution of fields that trace the mass
distribution of the large--scale structure.

Following \citet{scherrer} and \citet{scocci} we can write the two--point
function of the mass distribution as \bea
&&\bar{\rho}\xi({\bf x}-{\bf x}',z)=\nonumber \\
&&\int n(m,z) m^2 dm
\int d^3y u_m({\bf y},z) u_m({\bf y}+{\bf x}-{\bf x}',z) \nonumber \\
&+& \int n(m_1,z)m_1dm_1 \int n(m_2,z)m_2dm_2
\int d^3x_1u_{m_1}({\bf x}-{\bf x}_1,z) \nonumber \\
&\times& \int d^3x_2u_{m_2}({\bf x}'-{\bf x}_2,z)\xi({\bf
x}_1-{\bf x}_2;m_1,m_2,z), \eea where we have written out
explicitly the redshift--dependence of the correlation
function; hereafter, for simplicity, we usually drop this redshift--dependence 
when writing our expressions.  In above, $n(m)$ is the mass function
describing the number density of collapsed objects with given mass
$m$: \be \frac{m^2n(m)}{\bar{\rho}}\frac{dm}{m}=\nu
f(\nu)\frac{d\nu}{\nu} \ee where $\bar{\rho}$ is the background
density and $f(\nu)$ is defined, following arguments given by
Press \& Schechter (PS; 1974) as \be \nu f(\nu)=\frac{1}{2} \left(
\frac{\nu}{2\pi}\right)^{1/2} \exp{\left(-\frac{\nu}{2}\right)}.
\ee Here, $\nu$ is given by \be \nu\equiv
\frac{\delta^2_{\rm sc}(z)}{\sigma^2(m,z)}, \ee where $\delta^2_{sc}(z)$ is
the critical density for spherical clustering 
and $\sigma^2(m,z)$ is the density fluctuation
smoothed with a  tophat filter at a redshift of $z$ corresponding
to mass scale of $m$. Useful fitting formula for $\delta_{sc}(z)$ and related
variables can be found in \citet{Hen00}.
The first term of equation~(1) is the
two-point correlation function within one halo, the second term is
the two-point correlation function  between points being in two
different halos and $u_m({\bf x})$ represents mass distribution in
the collapsed region.

We apply the NFW \citep{nfw97} profile for the mass distribution in
the collapsed region with \bea
u_m(r)=\frac{\rho_s}{(r/r_s)(1+r/r_s)^2}, \eea where $r_s$ is the core
radius and $\rho_s$ is the density at $r_s$. We can also get $r_s$
from the known mean concentration relation,
$\bar{c}=r_{vir}/r_{s}$. We calculate the virial radius of each halo
following spherical collapse arguments such that $M = 4 \pi r_v^3
\Delta(z) \rho_b/3$, where $\Delta(z)$ is the overdensity of collapse
and $\rho_b$ is the background matter density today. We use comoving
coordinates throughout.

The mean concentration parameter
$\bar{c}$ is given by \citep{bullock,jing} \be
\bar{c}=\frac{9}{1+z}\left[\frac{m}{m_*(z)}\right]^{-0.13} \ee
where $m_*(z)$ is the critical mass at given $z$. The density
profile in Fourier space is \be u(k|m)=\int^{r_{vir}}_0 dr 4\pi
r^2 \frac{\sin kr}{kr} \frac{\rho(r|m)}{m}, \ee where $r_{vir}$ is
the virialized radius of the collapsed object.

The Fourier transformation of equation~(1) gives the power spectrum
\bea
P_{\rm dm}(k)&=&P_{\rm dm}^{\rm 1h}(k)+P_{\rm dm}^{\rm 2h}(k) \nonumber \\
P_{\rm dm}^{\rm 1h}(k)&=&
\int dm n(m)\left(\frac{m}{\bar{\rho}}\right)^2 u(k|m)^2 \nonumber \\
P_{\rm dm}^{\rm 2h}(k)&=&\int dm_1 n(m_1)
\frac{m_1}{\bar{\rho}} u(k|m_1)  \nonumber \\
&\times&\int dm_2 n(m_2)
\frac{m_2}{\bar{\rho}} u(k|m_2)  \nonumber \\
&\times&P_{\rm hh}(k|m_1,m_2). \eea Here, $P_{hh}(k|m_1,m_2)$ is the
halo--halo power spectrum which we take to be related to the power
spectrum of the linear density field, $P^L_{\rm dm}(k)$ by \be
P_{\rm hh}(k|m_1,m_2)=b_1(m_1)b_1(m_2)P^L_{\rm dm}(k). \ee 
The halo biasing $b_1$ is \citep{mo97} \be b_1(m)=1+\frac{\nu
-1}{\delta_{\rm sc}(z_1)}\,.\ee

\begin{figure}[t]
%\epsscale{0.2}

\plotone{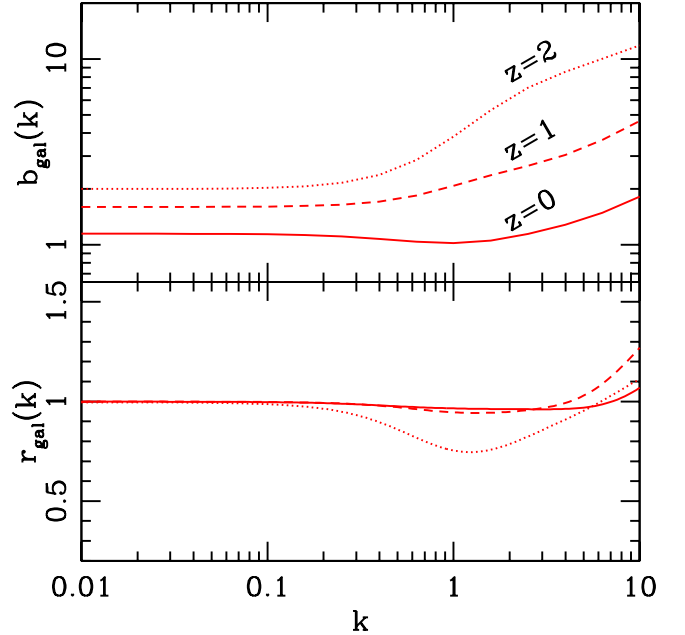}%[t]

%\vspace{-0.4cm}

\caption[]{\label{fig:bias}
Galaxy bias as a function of $k$ for our fiducial halo model at $z=0,1,2$
(upper panel) and galaxy--dark matter density contrast correlation coefficient
(lower panel).
}
\end{figure}

We assume that the dust giving rise to the FIRB is all in galaxies.
To describe clustering properties of these dusty galaxies, we modify
the halo model for dark matter to describe point sources following
previous discussions in the literature \citep{seljak,Jinetal98}.

An important ingredient is how these galaxies occupy dark matter
halos.  Following \citet{scocci,cooray02}, we model this halo
occupation number, the average number of galaxies in each dark matter
halo of mass $m$, through a simple relation of the form \be \langle
N_{\rm gal}(m)\rangle=A_{\rm gal}\left(m/m_0\right)^{p_{\rm gal}}. \ee
When illustrating our results, we will take $A_{\rm gal}=0.7,
m_0=4.2\times 10^{12}$ and $p_{\rm gal}=0.8$ consistent with what is
suggested in the literature for the halo occupation number of blue,
star-forming, galaxies \citep{scocci,cooray02}.  Besides these
parameters, we also set a lower bound on the halo mass, at $m_c =
10^{11} M_{\odot}/h$, in which dusty galaxies may exist. This accounts
for the fact that baryons may not collapse and cool to form stars in
low mass halos as well as to account for the fact that star--formation
may not be efficient due to processes such as supernova winds which
may expel any remaining gas in such small mass halos
\citep{benson,kauffmann}.

Note that we have described the halo occupation number of sources that contribute to
FIRB with a description that is consistent for blue galaxies at low redshifts. While these
late-type galaxies are expected to trace dusty-starbursts that form the FIRB, our choice of
parameters is also consistent with another consideration. The non-linear
clustering of sources in the IRAS Point Source Redshift Survey Catalog (PSCz), based on 60 $\mu$m fluxes, is well produced by a halo occupation number consistent with the above set of parameters for
$p$, $A_{\rm gal}$, and $m_0$ \citep{scocci,c02}. These sources are also likely to trace the
dusty-starbursts that contribute to FIRB. An additional complication, which we return to
later, is if FIRB sources are associated with either mergers or active galactic nuclei.

To describe galaxy clustering at the two-point level
we also need the second moment, $\langle
N_{\rm gal}(N_{\rm gal}-1)\rangle$ of the galaxy halo occupation
distribution. Given our limited knowledge, we take the simplest
approach involving a Poisson distribution such that $\langle
N_{\rm gal}(N_{\rm gal}-1)\rangle=\langle N_{\rm gal}(m)\rangle2$. This
assumption is consistent with numerical simulations especially at
the high mass end while at the low mass end of the mass function,
when $\langle N_{\rm gal}(m)\rangle <1$, evidence exists for departures
from a Poisson distribution \citep{scocci,cooray02}.

With information related to the halo occupation number, we can
write the power spectrum of dusty galaxies, again with two terms:
\bea
P_{\rm gal}(k)&=&P_{\rm gal}^{\rm 1h}(k)+P_{\rm gal}^{\rm 2h}(k) \nonumber \\
P_{\rm gal}^{\rm 1h}(k)&=&\int dm n(m)
\left(\frac{\langle N_{\rm gal}(m)\rangle}{\bar{n}_{\rm gal}(m)}\right)^2
u(k|m)^p \nonumber \\
P_{\rm gal}^{\rm 2h}(k)&=&\int dm_1 n(m_1)
\frac{\langle N_{\rm gal}(m_1)\rangle}{\bar{n}_{\rm gal}(m_1)}
u(k|m_1)  \nonumber\\
&\times&\int dm_2 n(m_2)
\frac{\langle N_{\rm gal}(m_2)\rangle}{\bar{n}_{\rm gal}(m_2)}
u(k|m_2)  \nonumber\\
&\times&P_{hh}(k|m_1,m_2). \eea Here, $p$ accounts for the
possibility that at least one galaxy may reside in the center of
each halo. Thus, we set  $p=1$ when $\langle N_{\rm gal}\rangle < 1$
and otherwise $p=2$. Here, $\bar{n}_{\rm gal}$ is the mean density of
FIRB sources given by 
\bea \bar{n}_{\rm gal}=\int dm n(m) \langle N_{\rm gal}(m) \rangle \, .
\eea 
The angular power spectrum of FIRB fluctuations can now be
obtained by projecting the three--dimensional dusty galaxy power
spectrum along the line of sight following standard techniques
such as the Limber approximation. The cross--correlation between
lensing and the FIRB, however, depends on the correlation of dusty
galaxy and dark matter distributions. The halo model provides a
simple calculation of this cross-correlation and we write \bea
P_{dm-gal}(k)&=&P_{dm-gal}^{\rm 1h}(k)+P_{dm-gal}^{\rm 2h}(k) \nonumber \\
P_{dm-gal}^{\rm 1h}(k)&=&\int dm n(m)
\frac{m}{\bar{\rho}}
\frac{\langle N_{\rm gal}(m)\rangle}{\bar{n}_{\rm gal}(m)}
u(k|m)^p \nonumber \\
P_{dm-gal}^{\rm 2h}(k)&=&\int dm_1 n(m_1)
\frac{m_1}{\bar{\rho}}
u(k|m_1)  \nonumber \\
&\times&\int dm_2 n(m_2)
\frac{\langle N_{\rm gal}(m_2)\rangle}{\bar{n}_{\rm gal}(m_2)}
u(k|m_2) \nonumber \\
&\times& P_{hh}(k|m_1,m_2).
\eea
The halo approach can be used to estimate the
dusty galaxy bias, $b_{\rm gal}$, and correlation, $r_{\rm gal}$, both as a
function of scale, $k$, and redshift $z$.
We define these two quantities as
\begin{eqnarray}
b_{\rm gal}(k,z)&=&\sqrt{\frac{P_{\rm gal}(k,z)}{P_{\rm dm}(k,z)}}\nonumber \\
r_{\rm gal}(k,z)&=&\frac{P_{dm-gal}}{\sqrt{P_{\rm gal}P_{\rm gal}}} \, ,
\end{eqnarray}
and plot them in Fig.1 as a function of $k$ for three redshifts
($z=$ 0, 1 and 2). Note that $b_{\rm gal}(k)$ is constant at the
large spatial scales corresponding to the linear regime. This
large--scale bias, under the halo model, is simply
\begin{equation}
b_{\rm gal}(k) = \int dm n(m) b_1(m) \frac{\langle N_{\rm gal}(m)\rangle}{\bar{n}} \, .
\end{equation}
At non-linear scales bias increases due to the choice of $p=1$ when
$\langle N_{\rm gal} \rangle < 1$.
The behavior of $r_{\rm gal}(k)$ can also be understood from 
the halo model. At linear scales,
galaxies trace the dark matter such that they perfectly correlate
with each other. At non-linear scales, $r_{\rm gal}(k)$ is greater
than unity since the galaxy power spectrum is defined such that,
following the standard definitions, it does not include the
shot--noise part associated with the finite number of galaxies
\citep{seljak}.

\section{Angular Power Spectra}

\begin{figure}[t]
%\epsscale{0.2}

\plotone{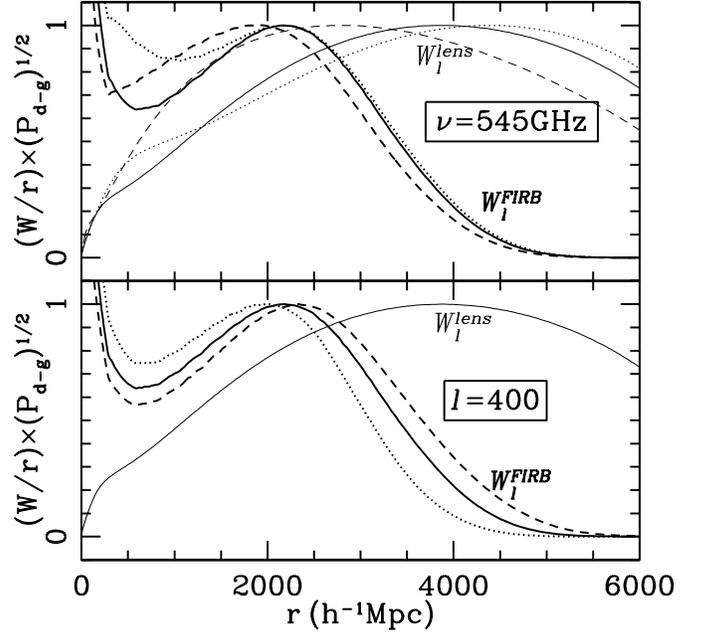}
%\vspace{-0.4cm}

\caption[]{\label{fig:weights}
Integrand of equation (\ref{eqn:cross-corr}).  Here
we plot the integrand of equation (\ref{eqn:cross-corr})
broken up into two factors, $W^{len}(k,\nu,z)/r\sqrt{P_{dm-gal}(l/r)}$ (light)
and $W^F(k,\nu,z)/r\sqrt{P_{dm-gal}(l/r)}$ (heavy).  Top panel
shows $l=100$ (dashed), $l=400$ (solid) and $l=1000$ (dotted) at $\nu = 545$ GHz
and bottom panel shows $\nu = 353$ GHz (dashed), $\nu = 545$ GHz (solid)
and $\nu = 852$ GHz (dotted) at $l=400$.
}
\end{figure}

In this paper we are interested in measuring the cross-correlation between
FIRB fluctuations and the lensing effect on CMB anisotropies. 
For the purpose of this discussion, we will describe lensing through the associated projected
potential, $\phi$, along the line of sight
\bea
\phi(\bn) = - 2 \int d\rad \frac{\da(\rad_0-\rad)}{\da(\rad)\da(\rad_0)}
                \Phi (\rad,\hat{{\bf m}}\rad ) \,,
\eea
where $\rad_0\equiv r(z=1100)$ corresponds to the radial distance to background CMB fluctuations which act as
a source for lensing and $\Phi$ is the gravitational potential field.
Here, $\da$ is the comoving angular diameter distance and, in a spatially-flat
cosmology, $\da \rightarrow \rad$. Note that the deflection angle of a CMB photon
is given by the gradient of this projected potential, $\alpha(\bn)=\nabla \phi(\bn)$.

First, we will consider the three power spectra associated with the
projected lensing potential, FIRB fluctuations at frequency $\nu$,
$F_\nu$, and the cross-correlation between the two. We use flat-sky
coordinates throughout the paper and write the Fourier transform of
quantities as \bea X(\vecl) = \int d^2\theta X(\theta) e^{-i \vecl
\cdot {\bf \theta}} \, , \eea where $X\equiv F,\phi$ for FIRB and
lensing potentials respectively.  We define power spectra of these
quantities and the cross-power spectrum such that \bea \langle
F_\nu(\vecl) F_\nu(\vecl') \rangle &=& (2\pi)^2 \delta_D(\vecl+\vecl')
C^{\rm FF}_{l}(\nu), \\ \langle \phi(\vecl) \phi(\vecl') \rangle &=&
(2\pi)^2 \delta_D(\vecl+\vecl') C^{\phi\phi}_{l}, \\ \langle
F_\nu(\vecl) \phi(\vecl') \rangle &=& (2\pi)^2 \delta_D(\vecl+\vecl')
C^{\rm F \phi}_{l}(\nu) \, . \\ \eea We can
write Fourier coefficients for fluctuations in either the FIRB or lensing
potential as \bea X(\vecl) = \int\frac{d^3k}{(2\pi)^3} \int d^2\theta
\int dr W^X(k,r) \delta_X(\vec{k},r) e^{-i (\vecl - \veck r) \cdot {\bf \theta}} \, .  \nonumber \\ \eea 
Following \citet{knox01}, we write the
FIRB weighting function at a given frequency $\nu$ as 
\bea
W^F(z;\nu) &\equiv& a(z)\bar{j}(\nu,z) /(2k_B)(c/\nu)^2 \,\, {\rm and}\nonumber \\
\delta_F &=& \delta_{\rm gal}.
\label{eqn:wfirb}
\eea
Similarly, for lensing potentials,
\bea 
W^\phi(k,z) &\equiv& -3 \frac{\Omega_m}{a(r)}
\left(\frac{H_0}{k}\right)^2\frac{\da(r_0-r)}{\da(r)\da(r_0)} \, \, {\rm and} \nonumber \\
\delta_\phi &=& \delta_{\rm dm}.
\label{eqn:wlens}
\eea
We have used Poisson's equation to replace the potential field with
the dark matter density field, resulting in the $k$--dependence of
the window function.

As written in equations~(\ref{eqn:wfirb}) and (\ref{eqn:wlens}), and discussed earlier, FIRB fluctuations
trace that of contributing galaxies, while 
lensing traces fluctuations in the dark matter field. 
For illustration purposes, we plot  both  FIRB and lensing weight functions in Fig.~\ref{fig:weights}
 as a function of multipole, or angular scale, and frequency. 
Note that the lensing weight function, while
peaking at nearly half the angular diameter distance to background source, is a broad function in redshift space.
This allows lensing potentials
to be correlated with a wide variety of large--scale structure tracer fields.
The main result from this comparison is that the lensing and FIRB weight functions overlap with each
other over a wide range in radial distance space. As we find later, this substantial overlap results
in a significant correlation between lensing potentials and FIRB fluctuations.

Using the weighting functions in radial space, we can now calculate the angular power
spectra of FIRB and lensing potential fluctuations as well as the cross angular power spectrum between the
FIRB and lensing potentials. We make use of the the Limber approximation \citep{lim,kaiser92} 
and write \bea C_l^{\phi \phi} &=& \int\frac{dr}{\da^2}[W^{\rm len}(k,z)]^2
P_{\rm dm}\left[k=\frac{l}{d_A},z\right] \, ,\nonumber \\
C_l^{\rm FF}(\nu)&=& \int\frac{dr}{\da^2}[W^F(z;\nu)]^2 P_{\rm gal}\left[k=\frac{l}{d_A},z\right]
\, , \nonumber \\ \label{eqn:cross-corr} 
C_l^{\rm F\phi}(\nu) &=& \int\frac{dr}{\da^2}W^F(z;\nu) W^{\rm len}(k,z)
P_{\rm dm-gal}\left[k=\frac{l}{d_A},z\right]\, , \nonumber \\
\eea
for lensing--lensing, FIRB--FIRB and FIRB-lensing power spectra.

\begin{figure}[t]
%\epsscale{0.2}

\plotone{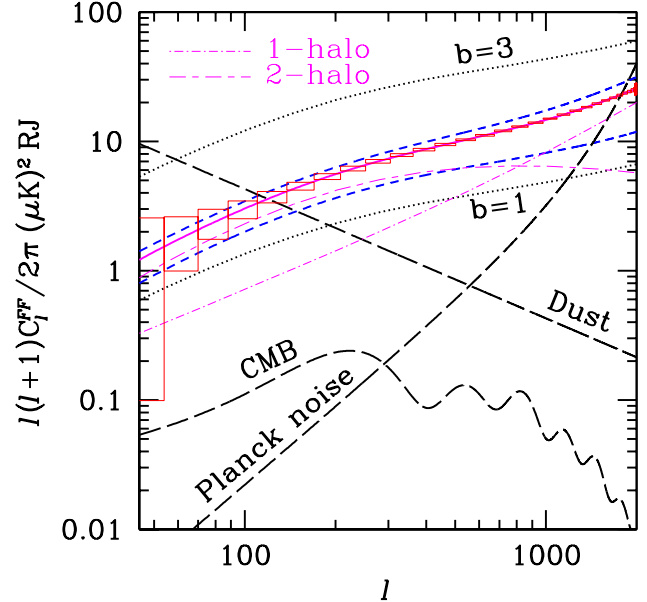}
%\vspace{-0.4cm}

\caption[]{\label{fig:clfirb} Angular power spectra at 545 GHz of
our fiducial FIRB model (solid with error boxes), CMB
(dot--dashed), galactic dust averaged over cleanest 10\% of the
sky (short/long dash) and Planck noise (dashed).  
The curves labeled 1-h and 2-h show the 1- and 2-halo contributions to the
FIRB power spectrum. Error
boxes are for measurement of FIRB power spectrum by Planck using
only the cleanest 10\% of the sky and including the contributions
from unsubtracted galactic dust.  The dashed lines below the
fiducial FIRB model are for angular power spectra for FIRB models
with $m_c \rightarrow 10^{10}M_{\odot}/h$ (lower) and $p_{\rm gal}
\rightarrow 0.9$ (upper).  Dotted lines are constant--bias FIRB
models with $b=1$ (lower) and $b=3$ (upper).}
\end{figure}

In Fig.~\ref{fig:clfirb} we show the angular power spectrum of the
FIRB for our fiducial model at 545 GHz. Here, we have calculated
five predictions for the FIRB signal including two constant bias
models that trace the non-linear dark matter power spectrum under
the formulation given by \citet{PeaDod96} with bias equal
to 1 (lower curve) and 3 (upper curve). The solid curve in the
middle is the prediction for the FIRB with our halo model. Because
$\nu = 545$ GHz is much greater than the frequency of peak
intensity for a 2.73 K black body, the CMB signal is significantly
smaller than the FIRB signal.

For comparison, in  Fig.~\ref{fig:clfirb}, we show the two contributions to the FIRB
power spectrum under the halo model involving 1- and 2-halo terms. As shown,
the FIRB clustering at tens of arcminute scales and above is described by correlations of
FIRB sources in different halos while at arcminute scales, the 1-halo term due to
correlations of sources within the same halo dominates. At these small scales, however, instrumental
noise starts to contribute rapidly contaminating the FIRB clustering at smallest scales.

\begin{figure}[t]
%\epsscale{0.2}

\plotone{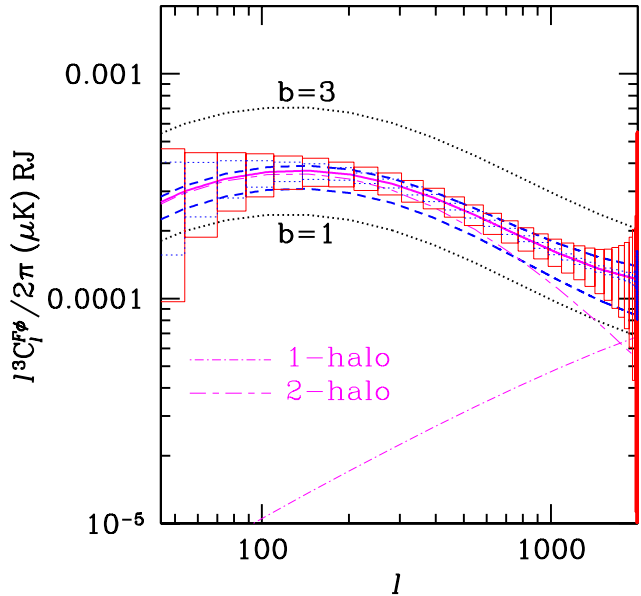}
%\vspace{-0.4cm}

\caption[]{\label{fig:blfl} Angular power spectrum of FIRB--CMB
lensing cross--correlation function at 545 GHz for same FIRB
models as in Fig.~\ref{fig:clfirb}.  The curves labeled 1-h and 2-h show the
1- and 2-halo contributions to the cross power spectrum. Note that in the range of
angular scales of interest to Planck, all contributions effectively come from the
2-halo term or the large scale correlations between FIRB sources and dark matter between different
halos. Error boxes assume use of
10\% of the sky, include the contributions from unsubtracted
dust and {\em do not} include polarization information. 
Solid (dashed) error boxes are for $\planck$ (no) noise. }
\end{figure}

\begin{figure}[t]
%\epsscale{0.2}

\plotone{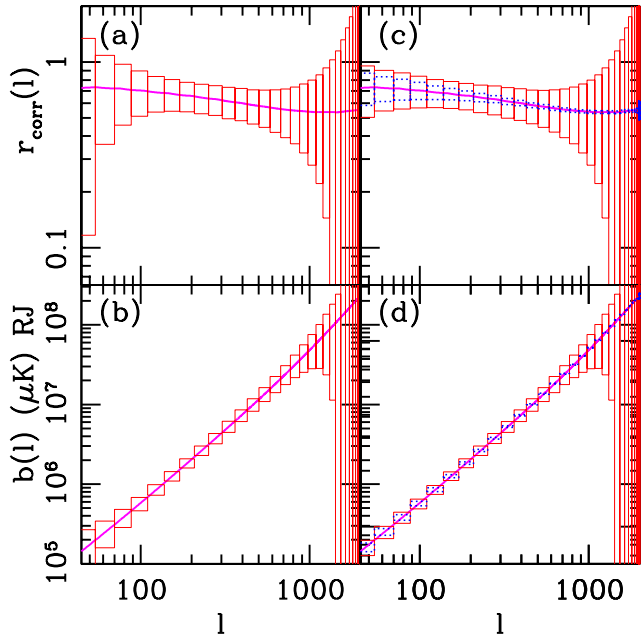}
%\vspace{-0.4cm}

\caption[]{\label{fig:rcoeff} {\it Top:} The correlation coefficient (see
equation (\ref{eqn:rcoeff})) at 545 GHz. (a) Error boxes assume 
the observation of the cleanest 10\% of the sky with $\planck$ noise
and include the contributions from unsubtracted dust.
(c) For comparison in this panel  we show the error boxes obtained when thereis no dust 
contribution to the FIRB map  (solid line) and, in addition, 
no noise contribution to the lensing reconstruction such that $N_l^{\phi\phi}=0$
(dashed line).
{\it Bottom:} The bias coefficient for FIRB fluctuations relative to that of the lensing
potential field. The error boxes follow those in the top panel.} 
\end{figure}

In Fig.~\ref{fig:blfl}, we show the cross-correlation between FIRB fluctuations at
545 GHz and the lensing potential. The curves shown here follow those
of Fig.~\ref{fig:clfirb}.  As shown, cross-correlation in the multipole range of interest
is dominated by the 2-halo term. This captures the large scale linear correlations between
FIRB sources and dark matter in different halos instead of the clustering of FIRB sources and
dark matter in the same halo. This dependence on the two-halo term is understandable from the
weight functions plotted in Fig.~1. As shown there, the cross-correlation peaks where the weight
functions overlap the most and this is at radial distance corresponding to redshifts of 1 and higher.
At such distances, the non-linearities, captured by the 1-halo term, do not dominate.

The strength of the correlation between $\phi$ and $F$ is
quantified by the cross--correlation coefficient \be
\label{eqn:rcoeff} r_{\rm corr}(l)=\frac{C_l^{\rm F\phi}}{\sqrt{C_l^{\phi\phi}C_l^{\rm FF}}} \,
. \ee In Fig.~\ref{fig:rcoeff} we show that the correlation is
moderate, as expected from the rough similarity of the weighting
functions.  Note that if fraction $1-\epsilon$ of the FIRB signal
is {\em completely} dependent on the lensing signal (i.e., can be
predicted from it) then we expect $1-r_{\rm corr} \simeq 0.5\epsilon^2$.  A
significant uncorrelated component still results in a $r_{\rm corr}$ near
unity; e.g., $\epsilon = 0.4 \rightarrow r_{\rm corr}=0.92$. 

In addition to cross-correlation, we can
also define a bias factor in Fourier space relating FIRB fluctuations and
dark matter traced by lensing potentials, such that 
\be b(l)=\sqrt{\frac{C_l^{\rm FF}}{C_l^{\phi\phi}}} \, . \ee
We also plot this bias factor and its associated errors on the bottom panel of Fig.~5.

\section{Error Forecasts}

Here we consider expected measurement errors on the various signals
predicted in the previous sections and shown in Figures 3, 4, and 5.  
First, we consider the expected errors on the detection of lensing potential power
spectrum and the FIRB power spectrum. Then we will discuss the associated errors on the
determination of the cross-correlation between FIRB and lensing.

\subsection{Lensing}

In the case of lensing power spectrum, we assume that the deflection angle associated with the
CMB lensing can be constructed from quadratic statistics involving temperature fluctuations
\citep{hu01,ck02}.

To understand the mechanism involved, first notice that 
lensing involves a remapping of the CMB temperature \citep{seljak96,zal00,hu01,ck02} such that
\begin{eqnarray}
\tilde \cmb(\bn) & = & \cmb[\bn + \nabla\len(\bn)] \nonumber\\
        & = &
\cmb(\bn) + \nabla_i \len(\bn) \nabla^i \cmb(\bn) \nonumber \\
&\quad& + {1 \over 2} \nabla_i \len(\bn) \nabla_j \len(\bn)
\nabla^{i}\nabla^{j} \cmb(\bn)
+ \ldots \,
\label{eqn:temp}
\end{eqnarray}
where $\cmb(\bn)$ is the unlensed primary  component of CMB, the contribution at the last 
scattering surface,
and  $\tilde \cmb(\bn)$ is the total contribution observed today after modifications due to
gravitational lensing deflections  during the transit.
To extract information related to lensing potentials, now 
consider a simple quadratic statistic involving the
squared temperature map, which in Fourier space can be written as
\begin{equation}
\hat{\cmb}^2(\vecl) = \int \frac{d^2\vecl_1}{(2\pi)^2}
W(\vecl, \vecl_1) \tilde \cmb(\vecl_1) \tilde \cmb(\vecl-\vecl_1) \, ,
\label{eqn:estimate}
\end{equation}
where $W$ is a filter that acts on the CMB temperature and can be derived by the requirment that
an ensemble average of 
this quadratic statistic returns an estimator for $\phi$, $\langle \hat{\cmb}^2(\vecl) \rangle
 \equiv \hat{\phi}(\vecl)$,
with a minimum noise contribution. Taking the Fourier transform of equation~(\ref{eqn:temp}),
and substituting for $\tilde \cmb(\vecl)$ in equation~(\ref{eqn:estimate}), one can obtain the
required filter as
\begin{equation}
W(\vecl, \vecl_1) = N_l^{\phi\phi}\frac{[\vecl \cdot \vecl_1 C_{l_1}^\cmb + \vecl \cdot (\vecl-\vecl_1) C_{|\vecl-\vecl_1|}^\cmb]}{2C_{l_1}^\tot C_{|\vecl-\vecl_1|}^\tot}\, ,
\end{equation}
with a normalization given by
\begin{equation}
[N_l^{\phi\phi}]^{-1} =  \int \frac{d^2\vecl_1}{(2\pi)^2} \frac{[\vecl \cdot \vecl_1 C_{l_1}^\cmb + \vecl \cdot (\vecl-\vecl_1) C_{|\vecl-\vecl_1|}^\cmb]^2}{2C_{l_1}^\tot C_{|\vecl-\vecl_1|}^\tot}\, .
\end{equation}
Here, $C_l^\cmb$ is the power spectrum of unlensed primordial temperature fluctuations, as measured at the last scattering
surface, while $C_l^{\tot}$ acounts for all contributions to the temperature power spectrum
such that $C_l^{\tot}=C_l^{\rm lensed-CMB}+C_l^{\rm noise}+C_l^{\rm fore}$. Here,
$C_l^{\rm lensed-CMB}$ is the CMB power spectrum after gravitational lensing, 
$C_l^{\rm noise}$ is detector and instrumental noise contributions, 
 and $C_l^{\rm fore}$ describes the contribution due to  any foregrounds and secondary effects, such as the
SZ effect.

Note that the variance of the filtered quadratic temperature map, 
\bea \langle \hat{\cmb}^2(\vecl) \hat{\cmb}^2(\vecl') \rangle = (2\pi)^2 \delta_D(\vecl+\vecl') \left[C_l^{\phi\phi}+N_l^{\phi\phi}\right] \, , \eea 
returns the lensing potential
power spectrum plus an associated Gaussian noise bias given by $N_l^{\phi\phi}$. 
We note that even in the case of a CMB experiment with no detector noise, the reconstructed potential, under the
quadratic statistic techhnique described above, is biased as 
$N_l^{\phi\phi}$ has contributions from primordial CMB fluctuations.  
This implies that the correlation coefficient defined in equation (\ref{eqn:rcoeff}) is not the correlation coefficient 
one measures between the FIRB map and the reconstructed $\phi$ map, as described above,
 even for an ideal experiment. This noise bias, however, can be eliminated possibly through a different reconstruction
technique involving an unbiased estimator of the lensing deflection potential. In order to generalize our
discussion, we ignore this noise-bias and treat $N_l^{\phi\phi}$ as the error contribution associated with
the reconstructed $C_l^{\phi\phi}$. Thus, correlation coefficients we plot in Fig.~5 are those that one would
measured by correlating the FIRB map with an unbiased estimator of the deflection angle.

In addition to the case with temperature we have just discussed, 
the polarization field can also be used to reconstruct $\phi$
\citep{gu00,benabed,hu01,ck02}, though we perform no error forecasts for polarization
data here. 

When estimating $N_l^{\phi\phi}$, we consider two experiments corresponding to
Planck and a no-detector noise, cosmic variance limited, experiment. In the case of
Planck, we use tabulated values for its detector sensitivities and resolutions, or beam sizes, following
\citet{Cooetal00} to calculate $C_l^{\rm noise}$. In the case of the cosmic variance
limited experiment, we set $C_l^{\rm noise}=0$ and take a maximum $l$ of 3000.
to which we integrate equation~(33). In both these experimental possibilities, we set $C_l^{\rm fore}=0$.

\subsection{FIRB}

On the FIRB side, we expect  a significant contribution to the angular power spectrum
at 545 GHz from galactic dust.  The level of the dust power spectrum
shown in Fig.~\ref{fig:clfirb} is for that in the cleanest 10\% of
$5^\circ \times 5^\circ$ patches \citep{knox01}.  The average
level in the cleanest 50\% is about 10 times higher.

We approximate the effect of dust contamination by including it as
an unsubtracted noise source with a power spectrum given by $C_l^{\rm D}$.
The error on the FIRB-FIRB power spectrum is
\bea \label{eqn:deltacl}
\Delta C_l^{\rm FF} & = &{1 \over \sqrt{(2l+1)f_{\rm sky}}}C_l^{\rm FF,tot} \, ,\nonumber \\
\eea
where following \citet{knox01} 
\bea
C_l^{\rm FF,tot}&=&\left(C_l^{\rm FF}+C_l^{\rm D}+N_l^{\rm FF}\right) \ {\rm and} \nonumber \\
N_l^{\rm FF} &\equiv &\Delta_{T_{\rm RJ}}^2 e^{l(l+1)\sigma2/8{\rm ln}2} \, .
\eea 
Here, $N_l^{\rm FF}$ is due to 
instrumental noise in the beam--deconvolved maps.
For example, the $\planck$ 545 GHz channel has $\sigma=5.0'$ and (on average)
$\Delta _{T_{RJ}}=13 \mu{\rm K-arcmin}$. Throughout we use the
$\nu = 545$ GHz channel as our probe of the FIRB though valuable, and complementary, information does lie in other channels at both higher and lower frequencies.

Equation (\ref{eqn:deltacl}) is formally correct for a
single--frequency measurement if the power spectrum of the dust is
perfectly known and the dust is a statistically isotropic Gaussian
random field.  These conditions do not hold; this equation only
serves as a rough guide, with dust increasing the errors when
$C_l^{\rm D} \ga C_l^{\rm FF}$ and having little effect when $C_l^{\rm D} <<
C_l^{\rm FF}$. Throughout we restrict our forecasting to results from
the cleanest 10\% of the sky, using the appropriate dust level.

\subsection{Lensing-FIRB}

Following our discussion on lensing reconstruction, one can use the estimator for the
deflection angle, or the projected potential, to measure the cross-correlation with FIRB fluctuations.
Thus, similar to equation~(34), we can write
$\langle \hat{\cmb}^2(\vecl) F(\vecl') \rangle = (2\pi)^2 \delta_D(\vecl+\vecl') 
C_l^{\rm F\phi}$  as an estimate of the cross-power spectrum between
lensing potentials and FIRB fluctuations.
In Fig.~\ref{fig:blfl} we plot $C_l^{\rm F\phi}$. Note that while $C_l^{\phi\phi}$ is biased
under quadratic temperature reconstruction techniques, the cross-correlation power spectrum
between lensing potential and FIRB is unbiased; this is due to the fact that FIRB fluctuations are assumed to be
uncorrelated with unlensed temperature anisotropies $\langle \cmb(\vecl) F(\vecl')\rangle=0$.

Following our discussion of errors on lensing and FIRB fluctuation power spectra, we can write
the expected error on $C_l^{F \phi}$ as
\bea
&&\Delta C_l^{\rm F\phi} = \sqrt{1\over(2l+1)f_{\rm sky}} 
\left[\left(C_l^{\rm F\phi}\right)^2+(C_{l}^{\phi\phi,tot})
(C_l^{\rm FF,tot})\right]^{1/2} \, ,\nonumber \\
\eea where $C_{l}^{\phi\phi,tot}=C_l^{\phi\phi}+N_l^{\phi \phi}$ 
is the total contribution in the $\phi$ power spectrum.

The associated error on the correlation coefficient and the bias
factor can be calculated by propagating errors related to
$C_l^{\phi\phi}$, $C_l^{\rm FF}$, and, $C_l^{\rm F\phi}$.  For
reference, we reproduce the final result for these errors here. In the case of
$r_{\rm corr}(l) \left(\equiv C_l^{F\phi}/\sqrt{C_l^{\phi\phi}C_l^{FF}}\right)$:
\begin{eqnarray}
&&\left(\Delta r\right)^2=\sum_l r^2_{\rm corr}(l) \Bigg\{
\left(\frac{ \Delta C_l^{\rm F\phi}}{C_l^{\rm F\phi}}\right)^2 + \frac{1}{ 2 (2l+1)f_{sky}} \times
\nonumber \\
&&\left[\left(\frac{C_l^{\phi\phi,\tot}}{C_l^{\phi\phi}}+
\frac{C_l^{\rm FF,\tot}}{C_l^{\rm FF}}\right)
\left(\frac{C_l^{\phi\phi,\tot}}{C_l^{\phi\phi}}+
\frac{C_l^{\rm FF,\tot}}{C_l^{\rm FF}}
- 4 C_l^{\rm F\phi}\right)\right]\Bigg\} \, ,\nonumber \\
\end{eqnarray}
while, in the case of $b(l) \left(\equiv \sqrt{C_l^{\rm FF}/C_l^{\phi\phi}}\right)$:
\begin{eqnarray}
&&\left(\Delta b\right)^2=\sum_l \frac{b^2(l)}{ 2 (2l+1)f_{sky}}
\left(\frac{C_l^{\phi\phi,\tot}}{C_l^{\phi\phi}}-
\frac{C_l^{\rm FF,\tot}}{C_l^{\rm FF}}\right)^2\, .\nonumber \\
\end{eqnarray}

\section{Discussion}

We now discuss the main results of our paper. 
We have introduced a semi-analytic approach to describe the  FIRB fluctuations.
The model involves a distribution of dark matter halos populated by dusty, starforming
galaxies. The basic ingredients of the calculation include properties of this
dark matter distribution, such as the mass function and clustering properties of dark matter 
halos, and properties of the sources responsible for the FIRB. 
To describe these sources 
we have introduced a relation, the halo occupation number,
 that attempts to capture how FIRB sources populate dark matter halos.
These model  can be used to calculate measurable properties of the sources 
such as their bias as a function of redshift and their spatial 
correlation with the underlying dark matter as a function of redshift.

The approach presented here differs from previous attempts
to model the fluctuations in the FIRB.  HK00 used several biasing
models, all of which were scale--independent.  The one most similar
to our calculation here assumed all the FIRB sources were in $10^{12} \msun$
halos and used the biasing prescription of \citet{mo96}.  Because
we include a range of halo masses extending below $10^{12} \msun$ and
because we take the FIRB mean to be $\sqrt{2}$ lower than HK00 did
our power spectra have smaller amplitudes.  The fluctuation power we
predict is also smaller than that in \citet{knox01} who used a constant
$b=3$ amplification of the non--linear power spectrum calculated using
the fitting formula of \citet{PeaDod96}. 

Even though we predict a smaller power spectrum than in the constant
 bias models we still expect FIRB fluctuations to be detectable in
 upcoming high frequency experiments such as Planck. Any
 multifrequency detection will allow the testing of our models and
 hopefully the extraction of interesting properties of our universe
 such as the evolution of the star--formation rate.  We do not discuss
 such possibilities in detail here as they have been already addressed
 in \citet{knox01} and references therein.

In this paper we have gone beyond the FIRB fluctuation power and focused on the cross-correlation between FIRB sources and the dark matter as traced by the intervening  lensing potential that  deflected CMB photons propagating to us.
Note that previous studies have already suggested that experiments such as Planck will
detect the lensing potential power spectrum with considerable signal-to-noise ratios \citep{zs99,zal00,hu01,ck02}.
In addition to the separate detections of FIRB and projected lensing potential power we have suggested that one can also perform a combined
study to measure the cross-correlation between FIRB fluctuations and the lensing potential.

We have discussed an extension of the estimators suggested for lensing
reconstruction in CMB data. Using quadratic statistics one can define an estimator of the
deflection angle, which can then be directly correlated with a high frequency map where
FIRB fluctuations are expected to dominate. Such a direct approach avoids complications
associated with other methods for extracting the lensing-FIRB cross power spectrum such as the direct measure of the three point function. 
As discussed in the literature, the FIRB-lensing correlation leads to a three-point
correlation function, or a bispectrum in Fourier space \citep{GS,CH00}.
Note that measuring the full configuration dependence of this bispectrum is difficult and currently
limited by computational methods  and measurement techniques. While improvements are expected,
our suggested technique avoids these issues by combing the different three point function configurations  in a particular way. 

As shown in Fig.~4, we find a considerable correlation
between FIRB fluctuations and the lensing potential. As illustrated in 
Fig. \ref{fig:weights}, this is due to the broad overlap between the respective radial weighting functions. Such a high correlation coefficient also leads to the  conclusion that
upcoming CMB experiments will be able to detect the cross power spectrum between FIRB and lensing with high signal-to-noise.  For Planck, we expect the cumulative signal-to-noise
to be of order 40 in the $\nu=545$ GHz channel. 

The high signal to noise expected for the detection  of the cross correlation and of the 
FIRB and lensing power spectra suggests that we will be able to  test our 
underlying model for the clustering of FIRB sources. As we discussed  our approach needed 
two main ingredients, the bias of FIRB contributing sources which was calculated in the halo model, and the  star formation rate which is needed to obtain FIRB weighting function. In turn, the 
halo approach depends on how starforming dusty galaxies populate dark matter halos.
As shown in Fig. 3 and Fig. 4 we expect that different values of the model parameters can be distinguished with the level of noise expected for Planck. 

In Fig.~\ref{fig:rcoeff}, we summarize our results with respect to how well the
projected correlation coefficient, $r_{\rm corr}(l)$, and the bias factor, $b(l)$,
can be measured with upcoming Planck data. While Planck allows a reliable detection of
the correlation coefficient as well as bias out to a multipole of $\sim$ 1000, there
are further improvements one can hope to achieve in the post-Planck era.
As summarized in Fig.~\ref{fig:rcoeff}(c), in the case of the correlation coefficient,
the error is dominated by the uncertainty associated with the lensing reconstruction.
With improved data involving better angular resolution and noise one can hope to
reach the limiting case of a perfect experiment which we have demonstrated with
a dashed line Fig.~\ref{fig:rcoeff}(c).

A few words of caution are required at this point. Note that we have described the FIRB 
and its
fluctuations as coming only from dusty starburst galaxies. In addition to the stellar contribution, there is also some from dust associated with torii surrounding active galactic 
nuclei. While the fraction of AGN contribution at far-infrared wavelengths could be as high as
30\% to 40\% \citep{alm01,risaliti02}, we have limited information on their redshift evolution so we have not included them in our calculation. The inclusion of AGNs within the halo model is trivial, the 
problems arise when trying to calculate the weighting function. 

Another caveat comes from our simple  description of how submm sources 
populate halos. While we have described them through a  halo occupation number, the physics 
is likely to be more complicated especially if dusty star formation is associated with mergers.
With a merger rate proportional to $N_{\rm gal}^2(m)$, the effective occupation number will probably depend on
a higher power of mass than considered here. In such a scenario, we expect source bias to be
larger with an increase in clustering power compared to the results presented here.  
Such an increase should be  detectable and constrained using observational
data.

While these issues may complicate the direct interpretation of the observations,
a reliable detection of the FIRB-lensing correlation can allow one to introduce more 
sophisticated analysis techniques to try to understand  these complicating factors. 
For example, the two power spectra and the cross-power spectrum can be combined
and written as $C_l^{FF} = b2(l) C_l^{\phi\phi}$ and $C_l^{\rm F\phi}=b(l)r(l)C_l^{\phi\phi}$,
which can be inverted with appropriate techniques to obtain $r(k,z)$ and $b(k,z)$.
Such an inversion requires accurate information on the FIRB radial weight function which 
 can be obtained observationally through the redshift distribution of sources that contribute to FIRB.
While unresolved surveys such as Planck will not allow such studies, in the future, targeted
studies of resolved FIRB sources, over small but representative patches of sky may provide the necessary information.  Constraints on this emissivity as
a function of redshift from current data are discussed in \citet{gispert00}, \citet{chary01} and \citet{takeuchi01}.  
With an inversion of two-dimensional clustering to  three-dimensions 
one can eventually constrain various aspects of the halo model such as the
occupation number \citep{c02}. 
While previous studies have motivated the use of FIRB fluctuations alone to
understand an important aspect of the large scale structure and its evolution history, we
also suggest that cross-clustering aspects, such as between FIRB and dark matter as traced by
gravitational lensing of CMB also plays an important role.

\acknowledgments

We thank Z. Haiman for the use of emissivity functions, $\bar
j(\nu,z)$, that he calculated. AC is supported at Caltech by a senior
research fellowship from the Sherman Fairchild foundation and DOE.  LK
is supported by NASA Grant NAG5-11098. MZ
is supported by NSF grants AST 0098606 and PHY 0116590, and by the
David and Lucille Packard Foundation Fellowship for Science and
Engineering.  This research was supported in part by the NSF under
Grant No. PHY99-07949.

\end{document}